\documentclass[twocolumn,amssymb,prl,showpacs,superscriptaddress]{revtex4}
\usepackage{epsfig}% Include figure files
\usepackage{graphicx}% Include figure files
\newcommand{\eg}{\epsilon_\gamma}
\newcommand{\eo}{\epsilon _0}

\begin{document}
\draft
\title{  Fano versus Kondo Resonances   in a Multilevel
``  Semi-Open ``  Quantum Dot }

\author{Piotr Stefa\'nski}
\email{piotrs@ifmpan.poznan.pl}
 \affiliation{Institute of Molecular Physics, Polish Academy
of Sciences, ul.~M.~Smoluchowskiego~17, 60-179~Pozna\'{n}, Poland}
\author{Arturo Tagliacozzo}
\affiliation{Coherentia-INFM, Unit\'a di Napoli and Dipartimento
di Scienze Fisiche, Universit\'a di Napoli "Federico II", Monte S.
Angelo-via Cintia, I-80126 Napoli, Italy}
\author{Bogdan R. Bu{\l}ka}
\affiliation{Institute of Molecular Physics, Polish Academy of
Sciences, ul.~M.~Smoluchowskiego~17, 60-179~Pozna\'{n}, Poland}
\date{Received \hspace{5mm} }

\date{\today}

\begin{abstract}
Linear conductance across  a large  quantum dot
via    a single level $\eo$ with  large hybridization  to  the  contacts
 is strongly  sensitive to quasi-bound  states  localized in  the
dot and weakly  coupled  to   $\eo$. The conductance  oscillates with  the
  gate  voltage due to  interference of  the
Fano type.  At  low  temperature and Coulomb
blockade,    Kondo  correlations   damp  the oscillations
   on  an  extended  range  of  gate  voltage values, by
 freezing  the  occupancy  of  the  $\eo$ level itself. As a consequence,
 the  antiresonances  of Fano origin  are washed  out.  The
results are in good correspondence with experimental data for a
large quantum dot in the semi-open regime.
\end{abstract}
 \maketitle

%\section{Introduction}
 Much  effort  is being  devoted  recently  in  the search  for  controlled
 transmission across  semiconductor quantum  dots (QD). Such  an  achievement allows to operate
 the   QD  as  a  coherent electronic  gate for individual electrons
 which is promising for nanoelectronic device applications \cite{springer}.

Linear Kondo  conductance  in  small  quantum  dots  (QD) coupled
to  a single  conduction channel,   at Coulomb Blockade (CB),  has
been  extensively  studied   since it  was first measured
\cite{GG,aleiner}. In dots  of  larger  size,
 irregularities  in  the  shape  and  defects  in  the  2D geometry
 can give rise to a sequence  of  quasi-bound levels $\eg$, that can  be
  localized  at  separate  sites within  the dot area.
A  gate  voltage  $V_g $  can  shift them  across  the chemical
potential $\mu $ and  they   become   occupied,  thus    changing the
occupancy  of  the  dot as  a whole.
Single  electron  capacitance  spectroscopy  $(SECS)$ is  a perfect tool  to
  monitor hundreds  of  electron  additions ($N \to N+1 $ )  into  such
  levels of a  larger  dot  and,  for  large $N$,  the dot charging energy
  $E_C$ is  the  energy  spacing  of  the  peaks \cite{zhitenev}.

 Multiple  resonances  and  dips  in  the
 conductance  have   been  recently  observed
   in  semi-open  structures \cite{fuhner},  pointing  to a Fano-like
 mechanism  \cite{fano}, with characteristic resonance  shape,
  classified by the  value of  the so-called Fano  parameter $q$. However,  classical Fano effect
 usually  corresponds  to  an   antiresonance with  total  extinction
of  the   conductance, which fails  to  appear  in  the  data.

Antiresonances  have  also been  predicted in    molecular  wires
connected  to metallic  contacts \cite{emberly},  but  they   have
very  different
  origin. If the  transmitted  electrons pass through  molecules,  the
antiresonance   arises from  the interference among   molecular
states,  in  the  absence  of  a
  continuum background   of  transmitting modes  which is  the typical
ingredient  for  Fano  resonances.

   Up  to  now,
  only  small interacting QDs  were  considered
\cite{gores}  and  studied  theoretically when  an additional
background of  continuum  states is  present, activated in the
semi-open regime \cite{bulka}.

 In  this  letter  we  show
that a semi-open  multilevel QD   can  bridge  the two extreme
cases  quoted  above,  although a  full   antiresonance is  very  unlikely
  to  occur.
 Most  remarkably, the  correlations in the  dot at  low  temperature
influence the  shape  of  the  resonances   in a striking  way.

We  assume  that  the  semi-open  dot  has  just  one  channel
open  to  conduction which  can  be  a single   level of energy $ \eo$ ($0-$level ),
with  an  extended  wavefunction  localized  within   the  dot  and  strongly coupled
 to the  Left and  Right electrodes,  so  that  it acquires  a
broad  linewidth $\Gamma  = \Gamma_R + \Gamma_L  $. The  $ \eo$
level participates  to  the conduction,  provided  it is located
not  too  far  in  energy  w.r.to
  $\mu$ ( taken  as  the reference energy). $\eo$  can  have  same  origin  as   a
sequence  of   localized  levels $\eg $, capable  of  binding
electrons.
 $\eg $  arise  in  the  spectrum
together with      $\eo$, each  time  the extended wavefunction of
the $0-$ channel  is    unable  to
  adjust  adiabatically  to  the space  variations of  the  confining
 potential from  source  to  drain \cite{gurvitz}.  In  our  model,
 the $\eg$  quasi-bound    states hybridize  with the $0-$ level
  wavefunction,  but  not  with  the  contacts, via    a   small
 electron  hopping  $t_\gamma $, with
  $ t_\gamma /t_{\alpha} \approx 0.2\div 0.3 $, where
  $t_{\alpha}$    ($\alpha=L,R$) are
the  hopping  matrix  elements  between  the  resonant $0-$level
  and  the  $L/R$  electrodes.

Although conductance  is  not  quantized  within  the $0-$channel,
some  features  of  CB  persist. A  gate  voltage  controls  addition of
electrons   in  the $\gamma $  levels.  Dynamical  Coulomb  interactions
on  the  localized  $\gamma $ levels  are  expected  not  to  influence the
 conduction, if cotunneling  and  non-perturbative  tunneling  processes
 involving
the  $\gamma $  levels  are neglected.  This  amounts  to  assume  that
 the other  $\gamma $ electrons  are  frozen  when
   each  new addition  occurs, so  that  Hartee-Fock  corrections  can  be
effectively  included  in  the  Koopman's  energies  $\epsilon_{\gamma}$
corresponding  to  quasiparticle $d^\dagger _{\gamma\sigma}$ operators for
 the    electron  that  is  being  added  ( with  spin  component $\sigma $
) (see below).
   We  assume  that in  the  range of occupancies (i.e.
$N$ values ) of  interest,  the addition  energy
 $\Delta \epsilon \approx  E_C $  and   roughly   constant,  so  that
we take  an  uniform  spacing  of  the levels   $\epsilon_{\gamma}$,
     with    $\Gamma >> \Delta  \epsilon $.

  When  one $ \eg$  is   moved across
$\mu$, the  conduction  electron of  the  $0-$ channel  is scattered  by
the localized  state  and an  antiresonance  appears in  the transmission.
The  shape of the interference pattern  depends on the position of
  $\eo$ w.r.to  $\mu $, similarly to  the  case of the
Fano-like transmission, where  the Fano factor  $q$ is  determined
by  the weakly energy  dependent phase shift of the continuum
states \cite{fano,stone}. Here   the  role  of  the  background
continuum is  played  by  the resonance  $\Gamma$  of  the  $0-$
level, provided  it  is  broad enough. We  also  include    an
Anderson
  onsite  repulsive Coulomb
 interaction  $U$  on  the  $0-$ level which  affects  conduction
 in  the  open  channel   at  low temperature $T$, even  if
$\eo  $ is    below  $\mu$.
 The  ratio   $U/t_{\alpha} $  is  taken  to  be not
very large,  ranging between $0.6\div 0.9 $.
  If  $T$ decreases
  below   the Kondo  temperature  $T_K$,   the  nature  of the   resonance
$\Gamma$ is  strongly modified  close  to  $\mu$,  by acquiring  a many-body
   Kondo imprinting.

 No additional  background of  continuum
 states  is  assumed  here,   with  $\Gamma >  U >\Delta\epsilon $,
     at  difference  with
 the   small  dot case \cite{gores,bulka},
 where, together with  the  background, $\Delta\epsilon  >U>\Gamma$.

The  striking feature that  we find for $U \neq 0 $ is that,  for  $T< T_K $,
  the very  possibility of
direct bare hopping   between the $0-$level  and  the $\gamma $
levels gives  rise to indirect correlated  hopping   at  $\mu $,
weakly  dependent on  the position of   $\eo$  near  the chemical
 potential. Consequently,  the shape of the bunch  of $\gamma $
  resonances is  markedly affected by  the  strong electron correlations
on  a  rather  broad  range  of  energies around $\mu$
(see  Fig. 4).
Namely, a  competition takes place between the Kondo resonance,
which forces  stabilization of  the occupancy number of the $0-$
level  to  $n_{0\sigma } =  1/2 $,
 and the quantum  Fano interference which fosters large
fluctuations of the particle number.
  This constitutes
a new aspect of the Fano effect which  is specific of  large quantum dots and
provides  a new  manifestation  of  the Kondo  conductance. We  believe
that the  recent  results  of F\"uhner  et  al. \cite{fuhner} can  be
  interpreted  in  this  way.

According  to   our  model,  the multilevel QD is  described
by the Anderson  Hamiltonian:
\begin{widetext}
\begin{eqnarray}
\label{for1} H= H_{dot}  + \sum_{k\sigma ,\alpha=L,R}\epsilon_{k\alpha}
c_{k\alpha \sigma}^{\dagger }c_{k\alpha\sigma} \:\:\: , \nonumber\\
H_{dot} =
\sum_{\sigma}\epsilon_{0}a_{\sigma}^{\dagger}a_{\sigma}+
Un_{0\uparrow}n_{0\downarrow}
 +\sum_{\gamma\sigma}\epsilon_{\gamma}d_{\gamma\sigma}^{\dagger}
d_{\gamma\sigma} +
 \sum_{\gamma\sigma}t_{\gamma}\lbrack
 d_{\gamma\sigma}^{\dagger}a_{\sigma}+h.c.
 \rbrack+\sum_{k\sigma \alpha}t_{\alpha}\lbrack c_{k\alpha \sigma}^{\dagger}
a_{\sigma}+h.c \rbrack .
\end{eqnarray}
\end{widetext}
Small  changes  of the   gate  voltage
  $V_g $   shift   the levels   $\eo$ and   $\eg$'s
uniformly  with respect to $\mu$.

 Electrons coming from the electrodes  probe the energy
spectrum of the dot. The conductance depends on  the spectral
density of the QD,   $\rho _\sigma  $,  and on its hybridization
$\Gamma_{\alpha\sigma}(\omega)=\pi
t_{\alpha}^{2}\rho_{L(R)\sigma}(\omega)$ with  the  electrodes. The
linear conductance
 (i.e. in zero limit of the drain-source voltage )
  can be written, in  units  of  the  quantum  conductance  $2e^2 /\hbar $,
in the form \cite{meir}:
 \begin{eqnarray}
 \label{cond}
 \mathcal{G}=\sum_{\sigma}\int_{-\infty}^{\infty}\Gamma_{\sigma}(\epsilon)\left(-\frac{\partial f(\epsilon)}{\partial
 \epsilon}\right)\rho_{\sigma}(\epsilon)d\epsilon,
 \end{eqnarray}
where $f(\epsilon)$  is the Fermi distribution function and
$\Gamma_{\sigma}(\epsilon)=\Gamma_{L\sigma}(\epsilon)\Gamma_{R\sigma}
(\epsilon)/[\Gamma_{L\sigma}(\epsilon)
+\Gamma_{R\sigma}(\epsilon)]$. Symmetric coupling is further
assumed. $\rho _\sigma  $  is
calculated from the imaginary part  of the retarded Green function
$G_{0,\sigma}$:
\begin{equation}
\label{g0}
 G_{0,\sigma}(\omega , V_g )=\left [ \left \{ G_{0,\sigma}^o\right \}^{-1}
 - \Sigma ^K _\sigma   -  \sum  _\gamma  W_\gamma
 \right ] ^{-1} \!\!\!\!\!\!\!(\omega , V_g ) \:\:\: ,
\end{equation}
 where $G_{0,\sigma}^o $  is  the bare
Green function of the level  $\epsilon_{0}$, in the absence  of
  Kondo  correlations, as  derived  from
  the  equation  of  motion
 $(\tilde\epsilon_{0(\gamma)} =\epsilon_{0(\gamma)} -V_g )$ :
\begin{eqnarray}
G_{0,\sigma}^o(\omega , V_g)= \lbrack   \omega- \tilde\epsilon_{0}
-\sum_{k\alpha} \frac{t_\alpha ^2}{\omega-\epsilon_{k\alpha
\sigma}}\rbrack^{-1}
\nonumber\\
 \equiv [\omega-\tilde\epsilon_{0}+i \Gamma ]^{-1}.
\end{eqnarray}
 If the
density of states in  the electrodes  is  structureless and broad,  the
$\omega $-dependence  of  $\Gamma $  can  be  ignored.  We  also lump
the  small  additional shift of $\epsilon_{0}$ due  to  the
hybridization with the electrodes in  $\eo$  itself.
$\Sigma ^K_\sigma $ is  the  Kondo  selfenergy.

   In
 Eq.~(\ref{g0}),  the hopping  onto  the
 $\gamma$ levels  gives  rise  to the  effective  selfenergy  correction
  for  the  resonant  state, $W_\gamma =  t_\gamma [\omega - \tilde\eg]^{-1}
  t_\gamma $.
On  the  other  hand, the propagator  of the $\gamma-$ electron has a pole
  with a
  small  but  finite imaginary part $\Gamma _\gamma $,  due to indirect
interaction with the electrodes.

\begin{figure}[!]
\epsfxsize=0.4\textwidth \epsfbox{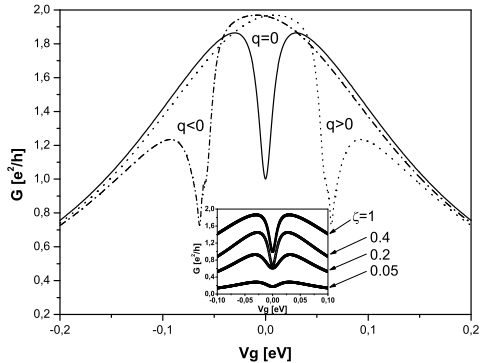}
\caption{\label{fig1}Conductance vs. gate voltage through QD for
$\gamma=1$ with $U=0$  at  $T=0$, showing  various Fano
$q$ parameter. In the inset,   the Fano dip for
$q=0$  appears  with  various values of $\zeta= \Gamma_L /\Gamma _R$.}
\end{figure}

In  Fig.~\ref{fig1} we plot  the  $T=0$ conductance  for  the case
$U  =0 $  ($\Sigma ^K = 0 $ )   vs. $V_g$,  when   $\eo  $
  is  located  right  at  the  chemical
potential ( $ \eo  = 0$ ), and  hybridization  to  the  leads gives  rise
  to a  broad resonance  of width  $ \Gamma  = 0.15 \: eV $.
The three thin  curves  show  the  conductance  of    just  one
single
  $\gamma $  level,   located  at
 $\eg = -0.05,0.0,0.05$, respectively.
 When  the  gate voltage $V_{g}$, is  such  that  the
 $\gamma$ level crosses the Fermi energy, a   resonance is
produced   with  the  characteristic  Fano-like  shape
corresponding to a Fano  parameter $q \: <0, \: =0, >0 $
respectively.  In   contrast  to  the Fano case,  however,  $
\mathcal{G}$  is  finite at the antiresonances.  It can be shown
that  for $q=0$ the  minimum reaches   $ \mathcal{G} = 1/2$ for
the most symmetric situation $\zeta = \Gamma _L /\Gamma _R =  1 $
( see inset of Fig.~\ref{fig1}). This difference is intimately
connected with a renormalization of $\gamma$ levels by indirect
coupling to electrodes. The finite imaginary part $\Gamma_\gamma$
does not cause full vanishing of the spectral density from
Eq.~(\ref{g0}) at the Fermi energy when the $\gamma$-level crosses
$\mu$. Thus, the  very  fact  that   the $\gamma$
levels  are  coupled only  indirectly to the electrodes  causes the
transmission at  the  antiresonances to  be  finite.
 As shown below, when $ U
\neq 0 $, no cancellation occurs as well.

 It  is  useful  to  write  down Dyson equation for $G_{0,\sigma}$
of  Eq.~(\ref{g0}), in  terms  of a
 $\bf{T}$-matrix  which  describes  the  correlated  tunnelling
 across  the device and to introduce Fano $q$ parameter
 \cite{stone,stefanski}. In the $U=0$ case we get
 $q(\omega=0,V_g)=-\tilde\epsilon_{0}/\Gamma$, so that
it increases linearly when
 $\epsilon_0$ is moved across Fermi energy.
 The phase shift acquired in  the  tunneling:
$\eta(\omega , V_g )=arg \: {\bf T} (\omega+i\delta , V_g)$ takes
the form in this case
$\eta(\omega=0,V_g)=-arctan[2\Gamma/\tilde\epsilon_{0}]$ when
a given  $\gamma$ level crosses the Fermi energy. Thus, we get a
relation between the phase shift and the Fano parameter:
\begin{equation}
\label{qfac}
 q _\sigma  ( V_g )  = 2 \cot{\eta _\sigma (\omega = 0 ,
V_g )}.
\end{equation}

The  occupation  number   $n_{0,\sigma }$ can  be  derived
directly from the  phase  shift  $\eta _{\sigma } (\omega ) $, by
means of the Friedel  sum  rule: $n_{0\sigma} =
\eta_{\sigma} (\omega  =  0  )/\pi$.  In   turn,  this relation,
together  with the expression for $q$, Eq.~(\ref{qfac})  allows
for a straightforward  interpretation  of the
 Fano parameter in  terms  of   $ n_{0\sigma}$.

\begin{figure}[!]
\epsfxsize=0.4\textwidth \epsfbox{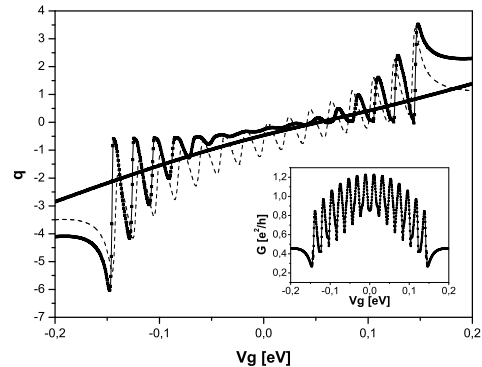} \caption{\label{fig2}
Fano $q$ vs. gate voltage $V_g$
 for  $\gamma=15$ levels :  $U = 0.1 eV$ ( oscillating full  curve),
 $U=0$ ( dashed line).The thick line is calculated for $U =0.1 eV$ in  the
  absence  of interference  due  to  the  $\gamma $ levels.
 The inset shows the  conductance  for  $\gamma=15$ and  $U=0$.}
\end{figure}

The wiggly transmission curve  in the inset of  Fig.~\ref{fig2}
corresponds  to  the  case of  a  bunch  of  15  resonances for
$U=0$ and $\Gamma=0.15 eV$. The  corresponding  dependence of  the
$q$ parameter of Eq.~(\ref{qfac})  on $V_g $ is presented in
Fig.~\ref{fig2}  by the dashed line. The oscillating  structure is
due  to  strong interference with the $\gamma$ levels.

 The  magnitude of  the  oscillations   is  reduced  when
$\epsilon_{0}$ approaches $\mu $ ($V_{g}\sim  0$). It
indicates the increased preference of electrons to resonate
across   $\epsilon_{0}$, instead  of  dwelling  in  the  dot   on
one  of the
 $\gamma$-levels. This tendency is further
enhanced when  $U \neq 0  $ and the  Kondo regime  sets  in (see below).

\begin{figure}[!]
\epsfxsize=0.4\textwidth \epsfbox{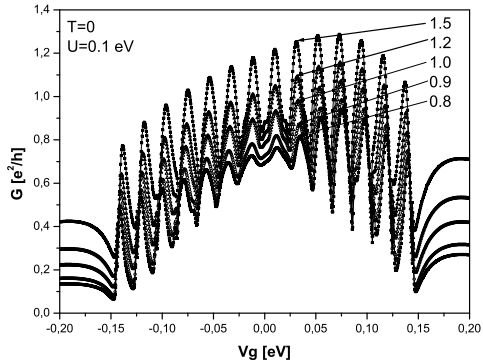} \caption{\label{fig3}
Conductance vs. gate voltage through the QD for
$U=0.1$ eV   at  $T=0$  for different ratios of
$(\Gamma_{L}+\Gamma_{R})/U$. See also the corresponding $T_K$
values in the inset of Fig.~\ref{fig4}.}
\end{figure}

Let  us  now  switch  on the  electron-electron  interaction  $U$
($U=0.1 eV$).
 We  include  electron-electron
interactions  within  the Interpolative Perturbative Scheme (IPS)
\cite{yeyati}.
 The  IPS calculated selfenergy  $\Sigma^{K}$ interpolates
between  two  correct limits:
 $ a )$  for  $U \to 0$: $\Sigma^{K}$  is derived
from    selfconsistent second order perturbation theory   in $U$
\cite{horv},
 $b) $ for  $\Gamma \to 0$: $\Sigma^{K} $ approaches
the  selfenergy  of  the  isolated impurity  level.
A selfconsistently calculated dynamical interaction  acting on
the  impurity level
ensures the fulfilment of the Friedel-Langreth sum rule.
 The  approximation has  been  found  to  give reliable  results
for a broad temperature range and up to  $U/\Gamma \sim 2.5
$\cite{horv}. Thus, it is especially suitable in the present
problem because  $U/\Gamma$  is  not  large. The conductance is
shown  in  Fig.~\ref{fig3} vs. $V_g$, for  various values  of
$\Gamma /U $  and  $T =0 $.
 When   $\Gamma/U > 1 $,   the  pattern does not look very different
  from  the  one  with  $U = 0 $.
The wiggles   are  superimposed on a   background which  is
  non symmetric, due  to  the  fact that
   the hump  in the  conductance is now shifted to
the value of $V_{g}$, for which $\epsilon_{0}=-U/2$.  There  the
conductance  attains  its  maximum.
In  any  other respect, the  Kondo resonance  is obscured
 in  the  spectral  density  by the  large  single-particle
broadening  $\Gamma $.

When   $\Gamma/U$ decreases, electron
correlations  prevail in the dot  and  show  up  in  the  conductance,
 as  seen  in  Fig.~\ref{fig3}.
An increasing  decoupling
of the QD from the electrodes causes the linewidth $\Gamma$  to  be
  strongly influenced  by  many-body  correlations, although  its  overall
width  is  not  sizeably  changed.

 The  signature   for  increasing Kondo  correlations is  the  heavy
  damping  of  the  oscillations  close  to  the  Fermi  energy
in  Fig. 3.
 This  corresponds  to  a
freezing  of  the occupation  number $n_{0,\sigma } $, for  values
of $V_g$  which  give  the  maximum  of  the   resonance as  shown
 in Fig.~\ref{fig2} by the wiggly heavy  line. Its  voltage  dependence
has  to  be   compared to
the one when  the $\gamma $ levels are  absent (straight heavy
line ).
 This  shows  that  there  is  an  extended
range  of energy  values
   at $\mu $ where  $ q $  is  pinned  at  the  value  $q  =0 $,
as  can  be seen  from  the shape  of  the  Fano-like  resonances.

\begin{figure}[!]
\epsfxsize=0.4\textwidth \epsfbox{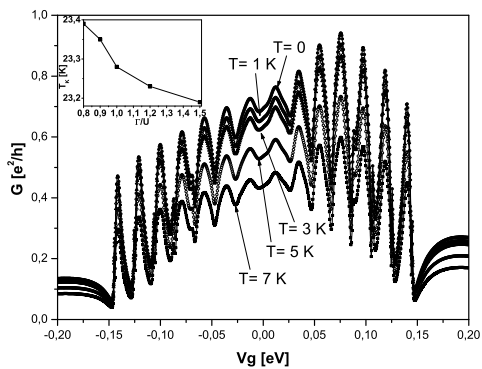} \caption{\label{fig4}
Conductance vs. gate voltage through the QD  for $U=0.1$ eV
at  various   temperatures. The
inset shows the dependence  of $T_K$ on the  ratio  $\Gamma/U$ for the
symmetric case ($V_g=0.05 eV$), as obtained from the
fitting to the Fermi liquid relation (see text).}
\end{figure}

In Fig.~\ref{fig4} conductance curves are calculated for various
temperatures    at  $\Gamma/U=0.8$. Rising the temperature, the
resonant tunneling is reduced but the damping is maintained. Kondo
physics still influences the conductance and indicates large Kondo
temperature, characteristic of the  mixed-valence regime. One parameter
scaling shows that the $\mathcal{G} (T)$ curves for various $V_g$
all collapse on  a single curve with $T^2$ dependence on  a wide
range of  $T$ values,  thus offering an estimate  of  the Kondo
temperature.
 $T_K$ vs. $\Gamma/U$ is presented in the inset of
Fig.~\ref{fig4} for the symmetric case at $V_g=0.05 eV$ as
obtained from the Fermi liquid formula
$\mathcal{G}=\mathcal{G}(T=0)[1-c(T/T_K)^2]$. The fitted $c$
parameter is in the range of $5.0-5.1$ to  be compared to the
value $ c=\pi^4/16\simeq 6.1$ of the   symmetric Anderson model in
the  Kondo regime \cite{costi}.

In conclusion, we considered a large multi-level QD  at  CB, with  one
single  conduction channel  in  the semi-open  regime.
  The  dependence of  the conductance   vs.
gate voltage  gives  evidence for  the competition between
  the  resonance of the  "background"  conduction channel,
 which  includes    Kondo  correlations, and  the
 Fano interference induced  by  a  bunch  of  quasi-bound  states
 localized  in  the  dot.  Oscillations are  markedly
  damped on  a broad  range  of energies  in the vicinity of the chemical
  potential. As a result, Kondo  correlations  wash  out  the  Fano
antiresonances  at $q=0 $.  The direct relation
between the phase shift and the pattern  of the Fano resonance  allows for
an  interpretation of experimental data,   as  the  ones  obtained  by
 F\"uhner  et  al. \cite{fuhner} for  a  semi-open  dot.
\begin{acknowledgements}
We  acknowledge  fruitful  discussions  with  Rolf Haug.
\end{acknowledgements}

\end{document}